# Status of the Finite Temperature Electroweak Phase Transition on the Lattice [*]

K. Jansen [a][†]

[a]Deutsches Elektronen Synchroton, DESY, Notkestr. 85,
22603 Hamburg, Germany

I review the status of non-perturbative investigations of the finite temperature electroweak phase transition by means of lattice simulations.

## 1. Introduction

Since it has been suggested [1] that the baryon asymmetry of the universe could have been generated at the symmetry restoring [2] finite temperature electroweak phase transition, much effort has been invested to study this idea in detail. If the above scenario is realized, the question of how the observed baryon asymmetry has been generated could be answered within the framework of the minimal standard model of the electroweak interactions. It is therefore a natural first approach to use perturbation theory for studying this question. Indeed, as of today results up to the 2-loop level are available [3–5,24]. However, it soon has become clear that even with the incorporation of special techniques to cure the infrared problems occuring in the symmetry restored phase, the perturbative series show bad convergence properties at least for Higgs masses larger than about 50GeV [5]. As a consequence one may question the reliability of the results as obtained in perturbation theory. In addition, it was found that for Higgs masses $M_H \gtrsim 80$GeV the perturbative series break down. Therefore only a small window of possible Higgs masses is left since the present experimental lower value is $M_H > 64$GeV [6]. This window would even be closed if one takes theoretical lower bounds for the Higgs mass as a result from vacuum instability [7]. Nevertheless, it would be very important to know whether perturbation theory can provide a quantitative description of the electroweak phase transition up to Higgs masses of about 80GeV. A positive answer would give strong confidence that we understand the properties of the electroweak phase transition.

It is obvious that only non-perturbative methods like numerical simulations can shed light on the reliability of the results as obtained in perturbation theory. Since the heart of the infrared problems arising in perturbation theory is the SU(2)-Higgs part of the minimal standard model, one can in a first approximation neglect the U(1) sector and the fermions. In this setup the problem becomes well suited for a numerical study since simulations of the SU(2)-Higgs model on the lattice are straightforward. In particular one does not have to deal with numerical simulations of chiral fermions on the lattice, a problem that as of today has not been solved [8].

Out of the three conditions for baryogenesis as formulated by Sakharov [9], –baryon number and CP violation and out of equilibrium processes– lattice simulations can test the last condition. The reason is that a possible realization of non-equilibrium behaviour is a first order phase transition. Lattice simulations are able to answer the question of the nature and –in case of a not too weak first order phase transition– the strength of the finite temperature phase transition. Whether the right amount of CP violation is generated will not be addressed here.

Any baryon asymmetry that has been generated at the electroweak phase transition is washed

---

[*]Plenary talk given at the International Symposium on Lattice Field Theory, 11-15 July 1995, Melbourne, Australia
[†]e-mail: kjansen@desy.de

out with a rate

$$R \propto \exp\left\{-O(1)\frac{4\pi v(T^{ph})}{gT^{ph}}\right\} \quad (1)$$

where $g$ is the electroweak SU(2) gauge coupling, $T^{ph}$ is the physical temperature and $v$ is the scalar vacuum expectation value to be taken at temperatures $T^{ph} \lesssim T_c$ with $T_c$ the physical transition temperature of the electroweak phase transition. Although the coefficient in the exponent is not known exactly, semi-classical estimates for its value exist [10]. Taking these estimates and comparing the rate in eq.(1) to the expansion rate of the universe leads to the rather conservative bound

$$v(T^{ph})/T^{ph} \gtrsim 1 \quad (2)$$

in order that baryon asymmetry is not washed out [11]. This implies that the phase transition not only has to be of first order but that it has to be strong enough, with "strong enough" defined by relation (2).

The values of the Higgs mass that have been investigated so far in numerical simulations are less than $M_H = 70\text{GeV}$. The reason for the restriction to these small Higgs mass values is that the strength of the phase transition weakens rapidly with increasing Higgs mass. For larger Higgs masses it becomes very difficult to resolve the order of the phase transition. Therefore all numerical work that is presented in this review mainly serves the purpose to confront results obtained in perturbative computations with the ones determined non-perturbatively by means of lattice simulations. Earlier review articles concerning the lattice approach to the finite temperature electroweak phase transition were given by Shaposhnikov [12] and by Kajantie [13].

## 2. The SU(2)-Higgs model: action and phase diagram

As mentioned in the introduction, the focus of this review will be on the SU(2)-Higgs part of the minimal standard model and the fermions as well as the U(1) part are neglected. To perform computer simulations a lattice regularization is adopted. The standard lattice form of the SU(2)-Higgs model is given by the action

$$S[U,\varphi] = \beta \sum_{pl}\left(1 - \frac{1}{2}\text{Tr}\, U_{pl}\right)$$
$$+ \sum_{x}\left\{\frac{1}{2}\text{Tr}\,(\varphi_x^+\varphi_x) + \lambda\left[\frac{1}{2}\text{Tr}\,(\varphi_x^+\varphi_x) - 1\right]^2\right.$$
$$\left. - \kappa \sum_{\mu=1}^{4}\text{Tr}\,(\varphi_{x+\hat\mu}^+ U_{x\mu}\varphi_x)\right\} \,. \quad (3)$$

Here $U_{x\mu}$ denotes the SU(2) gauge link variable, $U_{pl}$ is the product of four $U$'s around a plaquette and $\varphi_x$ is a complex $2 \otimes 2$ matrix in isospin space describing the Higgs scalar field and satisfying $\varphi_x^+ = \tau_2 \varphi_x^T \tau_2$. Often the length of the Higgs scalar field, $\rho_x^2 = \frac{1}{2}\text{Tr}\varphi_x^\dagger\varphi_x$, will be used. The plaquette term assumes the form of the pure Yang Mills action in the continuum limit when the lattice spacing $a$ is sent to zero. The parameter $\beta$ is related in the usual way to the SU(2) gauge coupling $g$ by $\beta = 4/g^2$. The terms on the second line are the scalar bare potential with bare quartic coupling $\lambda$. The last term is the gauge invariant kinetic part of the scalar fields. It assumes the continuum expression of the gauge covariant derivative when $a \to 0$. The scalar hopping parameter $\kappa$ is related to the bare mass $m_0$ by $m_0^2 = (1 - 8\kappa - 2\lambda)/\kappa$ and the quartic coupling $\lambda$ to its continuum counterpart by $\lambda_0 = \lambda/4\kappa^2$. Tuning $\kappa$ is essentially equivalent to tuning the bare mass parameter. In the following, the lattice spacing $a$ will be set to one unless otherwise stated. To realize a finite temperature the time extension $L_t$ of the lattice with volume $V = L_t L_x L_y L_z$ is periodized and kept much smaller than the the spatial directions $L_{x,y,z}$ which are chosen large enough to mimic infinite volume lattices.

At zero temperature considerable effort has been expended on the SU(2)-Higgs model in numerical lattice simulations [14]. The situation at a non-zero temperature is quite different and much less is known there. The pioneering works of [15] and [16] concentrated on large Higgs masses, that is on large values of the bare quartic coupling. The symmetry restoration picture could be verified in these works. It was also shown that the lattice data are consistent with an analytical crossover behaviour of the electroweak phase transition at large values of $\lambda$.

The results of these studies give the following schematic picture of the phase diagram in the 3-dimensional parameter space given by the couplings $\beta$, $\kappa$ and the temperature $aT^{ph} = T = 1/L_t$ depicted in fig. 1 for a fixed $\lambda \propto O(1)$. The phase diagram should be understood as being obtained on lattices with infinite extensions in the spatial directions and finite time extensions such that the highest temperature is $T = 1$ corresponding to $L_t = 1$.

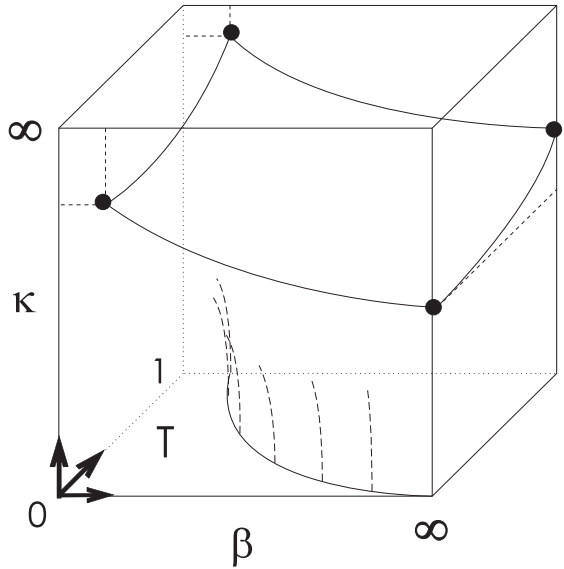

Figure 1. Schematic finite temperature phase diagram of the SU(2) Higgs model.

For $\kappa = 0$ we are left with the pure SU(2) gauge theory. This model exhibits at any $T > 0$ the finite temperature deconfinement phase transition at a critical value $\beta_c$ which shifts to larger $\beta$ values when $T$ is decreased. Switching on the scalar fields by taking a nonzero value of $\kappa$ this phase transition is stable and extents into the cube. For $\kappa$ large enough the transition may change to an analytical crossover behaviour as suggested in [16]. For $\beta = \infty$ the gauge fields are pure gauge and may be set to unity. The resulting model is the $\phi^4$ theory with global $SU(2)_L \otimes SU(2)_R$ or $O(4)$ symmetry. Keeping $T > 0$ fixed while varying $\kappa$ the electroweak finite temperature phase transition is passed at a critical $\kappa_c$. This phase transition separates the $O(4)$ symmetric from the symmetry broken phase with an residual $O(3)$ symmetry. Investigations of this model revealed only second order phase transitions [17–19]. The same is true if fermions with vector-like Yukawa interactions are added [20]. Releasing the temperature the finite temperature phase transition becomes a phase transition line. Choosing also $\beta < \infty$ this develops into a transition surface. For small $\beta$ one may, similar to the zero temperature case, expect a region in the parameter space where observables behave analytically and no phase transition takes place. For $\lambda$ and temperature large enough, also the electroweak phase transition may change into a crossover phenomenon. In summary there are three regions in the phase diagram:

- confinement region for $\beta < \beta_c$, $\kappa \ll 1$
- Higgs region for $\beta \gg 1$ and $\kappa > \kappa_c$
- deconfinement region for $\beta > \beta_c$, $\kappa < \kappa_c$

To make contact to continuum results like the ones obtained in perturbation theory, the lattice study has to control finite size effects and finite $a$ effects. Since a physical temperature is related to the time extent of the lattice by $T^{ph} = 1/L_t a$, the continuum limit $a \to 0$ corresponds to send the time extent $L_t \to \infty$. If one considers the case of finite temperature QCD [29] the necessary finite time extent of the lattice as used in the numerical simulations produce a severe limitation showing up in strong finite $a$ effects. This implies that large values of $L_t$ have to be taken. Although it will be shown below that the situation in the electroweak theory is different, this danger motivates the use of the dimensional reduction program that will be described in the next section.

Since the different parts of the phase diagram may be analytically connected it would be more appropriate to call them *regions*. However, in the literature –and sometimes in this article– they are also denoted as Higgs, confinement and deconfinement (or symmetric) *phases*. It is important to note that the finite temperature phase transition occurs at $\beta = 8 - 10$ if one takes the weak SU(2) gauge coupling as input. Therefore the transition occurs between the Higgs region and the deconfinement region. When $T$ is decreased to reach the continuum limit, while keeping the physical situation fixed, the finite temperature transition shifts to larger $\beta$-values. Thus the electroweak finite temperature phase transition always takes place between the Higgs and the deconfinement regions and does not interfere with the remnant of the pure gauge transition.

## 3. Dimensional reduction

In the euclidean path integral formulation of quantum field theories a finite temperature is introduced by periodizing in time. The Matsubara frequencies generated by this procedure can be separated into two kinds: The static Matsubara frequencies corresponding to massless zero modes and the non-static ones corresponding to massive modes. Since it is the zero Matsubara frequencies that lead to infrared problems in perturbation theory, it is a natural idea to integrate out all non-static modes. In this way the original 4-dimensional theory at finite temperature will be reduced to an effective 3-dimensional model. Of course, this will not cure the infrared problems associated with the static Matsubara frequencies. The idea in the approach of dimensional reduction is now to perform a simulation of the effective 3-dimensional model and treat the non-perturbative content numerically. The program as sketched above has been thoroughly discussed and carried out in [21-25] (for latest results, see [27,26]).

In practice the integration can not be performed exactly but the reduction is done in perturbation theory and one may wonder when this procedure is valid. In [25] this question has been investigated in detail. The reduction itself is justified when perturbation theory *at zero temperature* is valid. More specifically, the 4-dimensional finite temperature effective potential, $V_{eff}^{4d}$, is approximated by its 3-dimensional counterpart, $V_{eff}^{3d}$, through

$$\frac{V_{eff}^{4d} - T^{ph} V_{eff}^{3d}}{(T^{ph})^4} = O\left(\frac{m^2(T^{ph})}{(T^{ph})^2}\right) . \quad (4)$$

In (4) $m$ stands for a relevant mass scale in the zero temperature 4-dimensional theory. Since there all masses are proportional to the couplings $\lambda$ and $g$, dimensional reduction is valid when the 4-dimensional (renormalized) zero temperature couplings are small, i.e. when zero temperature perturbation theory works.

However, this does not imply high temperature perturbation theory to work. For this an additional condition has to be satisfied, namely that the ratio $g_3/m_T \equiv g^2 T^{ph}/m_T \ll 1$ with $g_3$ the 3-dimensional gauge coupling and $m_T$ a finite temperature mass scale. Performing the reduction perturbatively to 1-loop order one finds the following effective lattice action:

$$S = \beta_G \sum_{pl(3d)} \left(1 - \frac{1}{2}\mathrm{Tr} U_{pl(3d)}\right)$$
$$+ \frac{1}{2}\beta_G \sum_{x,i} \mathrm{Tr}\left\{A_0(x)U_i^{-1}(x)A_0(x+i)U_i(x)\right.$$
$$\left. - A_0^2(x)\right\} + \sum_x \left(5\Sigma - \frac{10}{3g^2\beta_G}\right) \mathrm{Tr} A_0^2(x)$$
$$+ \sum_x \frac{17 g^2 \beta_G}{48\pi^2}\left(\frac{1}{2}\mathrm{Tr} A_0^2(x)\right)^2$$
$$+ \frac{\beta_H}{2}\mathrm{Tr}\sum_{x,i}\left(\varphi^\dagger(x)\varphi(x) - \varphi^\dagger(x)U_i(x)\varphi(x+i)\right)$$
$$+ \sum_x \frac{1 - 2\beta_R - 3\beta_H}{2}\mathrm{Tr}\varphi^\dagger(x)\varphi(x)$$
$$- \frac{\beta_H}{2}\sum_x \frac{1}{2}\mathrm{Tr} A_0^2(x)\frac{1}{2}\mathrm{Tr}\varphi^\dagger(x)\varphi(x)$$
$$+ \beta_R \sum_x \left(\frac{1}{2}\mathrm{Tr}\varphi^\dagger(x)\varphi(x)\right)^2 . \quad (5)$$

In (5) the time component $A_0$ of the gauge field (in continuum notation) has been left in. In principle it can also be integrated out. The constant $\Sigma$ is given by $\Sigma = 0.252731$. Summations are to be taken over 3-dimensional lattices. The action contains the 4-dimensional gauge coupling $g$ and three additional bare couplings $\beta_G$, $\beta_H$ and $\beta_R$. They have a well defined relation to the 4-dimensional bare couplings. The coupling $\beta_G$ is given by

$$\beta_G = \frac{4}{g^2} \frac{1}{T^{ph} a} \ . \tag{6}$$

Eq.(6) implies that for fixed $g^2$ and physical temperature $T^{ph}$ tuning $\beta_G \to \infty$ corresponds to the continuum limit $a \to 0$. $\beta_R$ is given by

$$\beta_R = \frac{M_H^2}{8 M_W^2} \frac{\beta_H^2}{\beta_G} \tag{7}$$

with $M_H$ and $M_W$ the zero temperature 4-dimensional Higgs and W-mass, respectively. The parameter $\beta_H$ is left free and has to be tuned to the transition point.

As input 4-dimensional parameters, the gauge coupling $g = 2/3$, the W-boson mass $M_W = 80.6 \text{GeV}$ and as a free parameter the Higgs mass $M_H$, is chosen. Note that the Higgs mass is a renormalized $\overline{\text{MS}}$ mass to be taken at a scale $7 T^{ph}$. Its physical value might be different.

The numerical computations obey the following strategy. One first fixes $\beta_G$ which fixes the lattice spacing $a$. Then for a given lattice size $\beta_H$ and with the help of relation (7) $\beta_R$ are tuned to the transition point. The finite lattice transition points are determined by various methods (see [27] for a description of the methods applied). Although these methods may yield different $\beta_H^c$ on the finite lattice, the infinite volume limit extrapolation should give the same $\beta_H^c$. Therefore a careful finite size scaling is performed to obtain $\beta_H^c$ at infinite 3-volume $V_{3d}$. This procedure is repeated for larger and larger $\beta_G$ and hence for smaller and smaller $a$ values.

For every fixed set of parameters and with the non-perturbatively determined value of $\beta_H^c$ physical quantities can be computed. One example is the ratio $M_H/T_c$ which is given by

$$\frac{M_H^2}{4 T_c^2} = \left( \frac{g^2 \beta_G}{4} \right)^2 \left[ 3 - \frac{1}{\beta_H^c} + \frac{M_H^2}{4 M_W^2} \frac{\beta_H^c}{\beta_G} \right.$$
$$- \frac{9}{8 \pi \beta_G} \left( 1 + \frac{M_H^2}{3 M_W^2} \right) \Sigma - \frac{1}{2} \left( \frac{9}{4 \pi \beta_G} \right)^2$$
$$\left\{ \left( 1 + \frac{2 M_H^2}{9 M_W^2} - \frac{M_H^4}{27 M_W^4} \right) \log \frac{g^2 \beta_G}{2} \right. \tag{8}$$
$$+ \eta + \frac{2 M_H^2}{9 M_W^2} \bar{\eta} - \frac{M_H^4}{27 M_W^4} \tilde{\eta} \right\} \right]$$
$$+ \frac{g^2}{2} \left[ \frac{3}{16} + \frac{M_H^2}{16 M_W^2} + \frac{g^2}{16 \pi^2} \left( \frac{149}{96} + \frac{3 M_H^2}{32 M_W^2} \right) \right] \ .$$

In the above formula which is calculated in perturbation theory every quantity beside the constants $\eta$, $\bar{\eta}$ and $\tilde{\eta}$ is given. The constants can be fixed by relating the lattice scheme to a continuum scheme chosen to be $\overline{\text{MS}}$. In earlier works the constants have been evaluated by computing condensates like $< \frac{1}{2} \text{Tr} \varphi_x^\dagger \varphi_x >$ in the $\overline{\text{MS}}$ scheme and measuring it directly on the lattice. In the relation of the condensates computed in both schemes the unknown constants appear. Measuring $< \frac{1}{2} \text{Tr} \varphi_x^\dagger \varphi_x >$ at various coupling parameter values allows therefore for a fit from which the constants can be extracted.

Recently, Laine [28] performed a full 2-loop perturbative computation relating the lattice to the $\overline{\text{MS}}$ scheme. This allows for a complete 2-loop perturbative computation of the numerical values for the above mentioned $\eta$ constants. Comparing these 2-loop results with the earlier Monte Carlo results for the constants a satisfactory agreement was found. Taking eq.(8), an infinite volume –but still non zero lattice spacing– value for $M_H/T_c$ can be given. By repeating the numerical simulation at larger values of $\beta_G$, and therefore at smaller lattice spacings, finally the continuum limit extrapolation can be achieved.

The procedure is illustrated in fig. 2 and fig. 3. Fig. 2 shows the infinite volume extrapolation of $\beta_H^c$ at fixed $M_H = 35 \text{GeV}$ and fixed $\beta_G = 8$. The figure nicely demonstrates the convergence to the same infinite volume $\beta_H^c$ from all methods used. Fig. 3 shows the extrapolation to the continuum $a = 0$. Note the scale in the plot which indicates the very precise determination of $T_c$.

Progress since the Bielefeld conference last year in the 3-dimensional reduced model consists in two things. First, there is now a full 2-loop analytical expression of relating lattice data to their continuum counterparts in the $\overline{MS}$ scheme. Second, new simulation results have been obtained at $M_H = 60$ and $M_H = 70$ GeV. Difficulties to resolve the order of the phase transition at $M_H = 80 \text{GeV}$ were reported at this conference

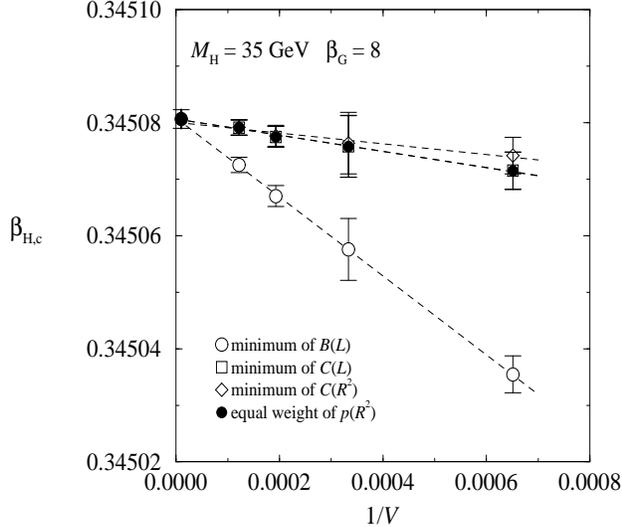

Figure 2. Infinite volume extrapolation for $\beta_H^c$ from four different methods as explained in [27].

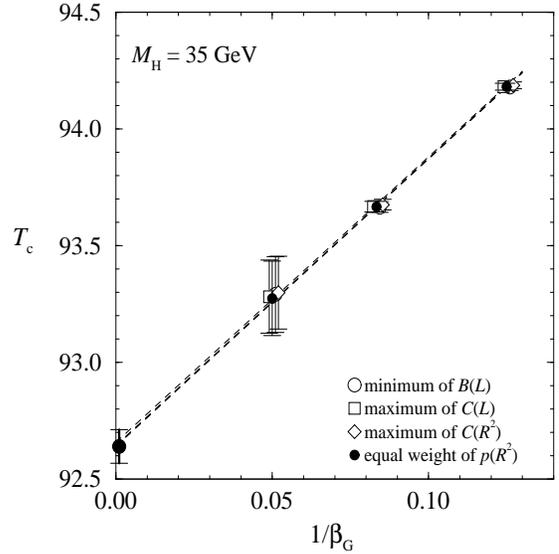

Figure 3. Continuum limit extrapolation of the critical temperature $T_c$ given in GeV.

[27]. In table 1 a comparison is made between the transition temperature as obtained from the simulations in the 3-dimensional reduced model and perturbation theory done in the effective model in 3 dimensions. As a second quantity for comparison the ratio $v(T_c)/T_c$ is taken. Note that the data in table 1 can not be directly related to 4 dimensional values.

Table 1 demonstrates the high precision that can be obtained in simulations of the SU(2)-Higgs model. Differences between the perturbatively obtained results and from the simulations can be seen. But these discrepancies are very small indicating also only small non-perturbative effects.

### 3.1. Critique on dimensional reduction

Recently, Jakovác [30] computed the 3- dimensional reduced theory up to 2-loop order in a scalar field theory. The effective model in three dimensions is super-renormalizable. All counter terms that are generated in the reduction step have to be cancelled by terms appearing in this super-renormalizable theory. Although this criterion works perfectly at the 1-loop level, a clear mismatch is encountered at 2-loop order. The counter terms generated in the reduction process can no longer be cancelled by the ones in the 3-dimensional theory, see also [31].

Cancellation can only be achieved again if non-local terms in the effective action are allowed. These non-local terms are expected on quite general grounds to appear on the scale $(2\pi T)^{-1}$ [32]. Splitting the low frequency from the high frequency modes and treating the low frequency modes as the block averages [33] one obtains an expression for the finite temperature propagator

$$\Gamma_T(r,t) = T \sum_{n \neq 0} \int \frac{d^3 p}{(2\pi)^3} e^{-ipr - i\omega_n t} \frac{1}{\omega_n^2 + p^2} \quad (9)$$

which leads to exponential decay modes with decay length $(2\pi T)^{-1}$. To circumvent this problem, in [30] it was suggested to integrate out all modes with $p \gg T$ and to obtain in this way an effective cut-off theory. A different approach was followed in [32]. There it was suggested to stay on the lattice and to perform block-spin transformations –either analytically or numerically– to separate the low from the high frequency parts. For finite temperature systems the block size in the temporal direction should be taken to be $1/T$. The remaining (perfect) effective block spin action is then purely 3-dimensional. Results for the elec-troweak phase transition emerging from this attractive concept have still to be awaited for. In a similar spirit is the average action approach in the continuum [34]. In the approach of dimensional reduction the problem of the non-local terms is treated by matching Green's function of the 3-dimensional and 4-dimensional theories [24,26].

Table 1
Comparison of the perturbatively [pert] and numerically [latt] obtained results in the 3-dimensional reduced model. $M_H$ and $T_c$ are given in GeV.

| $M_H$ | $T_c$ [pert] | $T_c$ [latt] | $v(T_c)/T_c$ [pert] | $v(T_c)/T_c$ [latt] |
|---|---|---|---|---|
| 35 | 93.3  | 92.64(7)  | 1.75 | 1.86(3)  |
| 60 | 140.1 | 138.20(4) | 0.68 | 0.691(7) |
| 70 | 157.0 | 154.5(1)  | 0.55 | 0.57(2)  |

## 4. Four dimensions

The conceptually cleanest and clearest way to treat the finite temperature SU(2) Higgs model is, of course, a direct simulation of the four dimensional theory at finite temperature. However, as Kanaya's talk at this conference again confirmed, the lesson from finite temperature QCD simulations is that one might have to expect sizable scaling violations. This would result in large values of $L_t$ to suppress finite $a$ effects which in turn necessitates large lattices since one wants to keep $L_t \ll L_s$. This danger motivated the dimensional reduction program. It seems, however, that we are fortunate in the SU(2)-Higgs model. As will be shown below, scaling violations for the quantities of interest are surprisingly small and safe extrapolations to continuum physics are possible.

Lattice simulations of the 4-dimensional theory have been performed by two groups. In [35] the finite temperature simulations at small $\lambda$ were initiated. Newer results of this group were presented last year at the Bielefeld lattice conference [36,37]. Results for a determination of static potentials can be found in [38]. A high statistics simulation of the 4-dimensional SU(2) Higgs model was performed in [39–42]. Due to the use of the Alenia Quadrics (APE) massively parallel machine and of algorithms specifically designed for the SU(2)-Higgs model [43,36] a high statistics and precise data could be obtained. It should be noted that the realization described in [36] of the method introduced in [43] gives the best results for the autocorrelation time. Fortunately enough, both groups performed simulations at the same parameter values. The results they obtain for various quantities are in good agreement. In the following I will therefore concentrate on the methods and results of only one group.

The basic goal of all the simulations is to make contact to continuum physics. To reach the continuum limit one has to run along renormalized trajectories also called lines of constant physics. Of course, the $a = 0$ limit can never be reached in the SU(2)-Higgs model because of triviality. What we mean by continuum limit therefore is to approach the scaling region of the SU(2) Higgs theory where cut-off effects are negligible. There the approximation of the continuum cut-off theory is expected to be very good in the low energy regime. The lines of constant physics are given by dimensionless ratios of physical quantities or the dimensionless coupling constants $g_R$, the renormalized gauge coupling and $\lambda_R = M_H^2/8v^2$, the renormalized quartic coupling. It is natural to follow perturbation theory and perform the renormalization at zero temperature.

In the 4-dimensional approach to the finite temperature electroweak phase transition in a first step the bare parameters of the action (3) are fixed such that the desired physical situation is realized. As physical input ones takes the W-boson mass $M_W = 80\text{GeV}$ and the gauge coupling $\beta = 8$ which amounts to $g^2 = 0.5$. Fixing the Higgs mass basically amounts to fixing the bare quartic coupling $\lambda$. Choosing $\lambda = 0.0001$ the Higgs mass will be around 20GeV and $\lambda = 0.0005$ corresponds to $M_H \approx 50\text{GeV}$. The physical temperature $T^{ph} = T/a$ is given by the temporal extent $L_t$ of the lattice. The phase transition itself is then found by tuning the scalar hopping parameter $\kappa$ to its transition value $\kappa_c$ (see fig. 1). At $\kappa_c$ one determines the order of the phase transition. If it is first order, quantities like surface tension, latent heat, jump of the order parameter etc. are computed. To get the values for the renormalized zero temperature quantities additional simulations at $\kappa = \kappa_c$ are needed but now at zero temperature which amounts to choose a hypercubic symmetric lattice which is large enough to suppress finite size effects. Note that $\kappa = \kappa_c$ at zero temperature belongs always to the symmetry broken Higgs phase, see fig. 1.

The approach to the continuum limit $a \to 0$ is realized by increasing $L_t$ while staying on the lines of constant physics. To compensate the change of $a$ the bare parameters $\beta$ and $\lambda$ have to be changed accordingly. To estimate the values of these parameters when going from $L_t$ to $L_t + 1$ the 1-loop continuum renormalization group equations are used

$$\frac{dg^2(\tau)}{d\tau} = -\frac{43}{48\pi^2} g^4 + \mathcal{O}(\lambda_0^3, \lambda_0^2 g^2, \lambda_0 g^4, g^6) \ ,$$

$$\frac{d\lambda_0(\tau)}{d\tau} = \frac{1}{16\pi^2}\left[96\lambda_0^2 + \frac{9}{32}g^4 - 9\lambda_0 g^2\right]$$
$$+ \mathcal{O}(\lambda_0^3, \lambda_0^2 g^2, \lambda_0 g^4, g^6) \qquad (10)$$

with $\tau = \ln(aM_W)^{-1}$.

From the above discussion it should become clear that a first important ingredient in the study of the electroweak phase transition is the determination of the transition $\kappa_c$. For this three methods have been used. It is noteworthy that in the 4-dimensional simulations for the determination of all relevant quantities at least two independent methods are used, providing very useful cross checks. The most precise method to determine $\kappa_c$ is the multicanonical simulation technique [44] which gives precision data for the distribution of suitable order parameters. Applying the equal height or equal weight criterion to these distributions allows to give $\kappa_c$ up to seven digits [42]. The presumably most practical method is the so called $2 - \kappa$ method [40,42]. Here one chooses elongated lattices and sets one half of the lattice to $\kappa_1$ belonging to the symmetric and the other half to $\kappa_2$ belonging to the Higgs region of the model. In this way an interface is generated between the two halfs. Bringing both $\kappa$ values closer to $\kappa_c$, eventually the interface will break down. The corresponding $\kappa_1$ and $\kappa_2$ give a lower and upper bound for the transition $\kappa_c$. This method works to a good and satisfactory precision and is mostly used for a first estimate of $\kappa_c$.

As an analytical tool the effective potential can be used. In [40] the gauge invariant effective potential [45,46] has been used. It was recently found that at least in 1-loop order the effective potential in Landau gauge is equivalent to the gauge invariant one if one chooses a particular mass resummation in the Higgs region by setting the Goldstone boson mass to zero [5]. One obtains for the broken phase

$$V_{1-loop} = V_{tree} \qquad (11)$$
$$+ \int_{-\pi}^{\pi} \frac{d^4k}{(2\pi)^4} \left\{\frac{9}{2}\ln(\hat{k}^2 + m_g^2) + \frac{1}{2}\ln(\hat{k}^2 + m_\phi^2)\right\} \ ,$$

where the masses are related to the parameters in the lattice action (3) and the translational invariant saddle point solution $\rho^2$ [5] by

$$m_g^2 = \frac{1}{2}\kappa g^2 \rho^2 \ , \qquad m_\phi^2 = \frac{4}{\kappa}\lambda\rho^2 \qquad (12)$$

and the momenta in the lattice integrals are $\hat{k}^2 = \sum_\mu [2 - 2\cos(k_\mu)]$. The tree level potential reads

$$V_{tree}(\rho^2) = (1 - 8\kappa)\rho^2 + \lambda(\rho^2 - 1)^2. \qquad (13)$$

The Landau gauge effective potential allows for a standard renormalization. One defines the renormalized mass $M_R^2/Z_R = d^2V/d\rho^2\big|_{\rho=\rho_{min}}$ and the renormalized coupling by $\lambda_R/Z_R^2 = d^4V/d\rho^4\big|_{\rho=\rho_{min}}$. Here $Z_R$ is the wavefunction renormalization constant [47]. The renormalized vacuum expectation value is $v_R = v/\sqrt{Z_R}$. For the value of the gauge coupling used in evaluating the effective potential the mean field improvement is taken, $g^2 \to g^2/<U_p>$ [48] with $<U_p>$ the averaged measured plaquette value. When below lattice simulation results are compared to lattice perturbation theory, the effective potential (11) and the above definitions of renormalized quantities are used. Note that in the effective potential approach for the renormalization step at zero temperature the value of $\kappa_c$ is taken as determined from the effective potential itself.

All three methods mentioned above give a first order phase transition. The most clean signal comes from the distributions of suitable observables using multicanonical simulation techniques. They show very pronounced two peak signals with about 10 orders of magnitude difference between the maximum and the minimum of the distribution [40]. After having identified the first order nature, as a second step computations of quantities characterizing the strength of the phase transition are necessary. A first such quantity is the latent heat.

### 4.1. Latent heat

Its calculation may proceed in the following way [5]. At the phase transition the free energy $W$ is continuous and

$$W_s(T, J(t)) = W_b(T, J(T)) \qquad (14)$$

determines the phase boundary. Here $W_{s(b)}$ is the free energy of the symmetric (broken) state. The external source $J$ is thought of being coupled gauge invariantly to the composite operator $\text{Tr}\varphi^\dagger\varphi$. Taking the total variation in eq.(14) one obtains

$$\frac{\partial}{\partial T}(W_s - W_b) = -\frac{\partial}{\partial J}(W_s - W_b)\frac{dJ(T)}{dT} \qquad (15)$$

which gives the latent heat $\Delta Q$ through $\frac{\partial}{\partial T}(W_s - W_b) = -\Delta Q/T$ and the order parameter jump $\Delta\rho^2$ via $-\frac{\partial}{\partial J}(W_s - W_b) = \Delta\rho^2$. As a result the well known Clausius-Clapeyron equation

$$\Delta Q = \Delta\rho^2 T \frac{dJ}{dT} \qquad (16)$$

is obtained.

For the model under consideration for dimensional reasons we have $m_0^2 + J(T) = C(g^2, \lambda)T^2$. And therefore

$$T\frac{dJ}{dT}\bigg|_{J=0} = 2m_0^2 \qquad (17)$$

and finally

$$\Delta Q = -M_H^2 \Delta\rho^2 (1 + O(g^2, \lambda)) \qquad (18)$$

with $M_H = \sqrt{-2m_0^2}$. This relation connects the order parameter jump to the latent heat. A similar expression was obtained and used also in the 3-dimensional reduced model. There is a second more complicated expression for the latent heat that can be obtained by differentiating the action density with respect to the lattice spacing. This second definition is explained and used in [40].

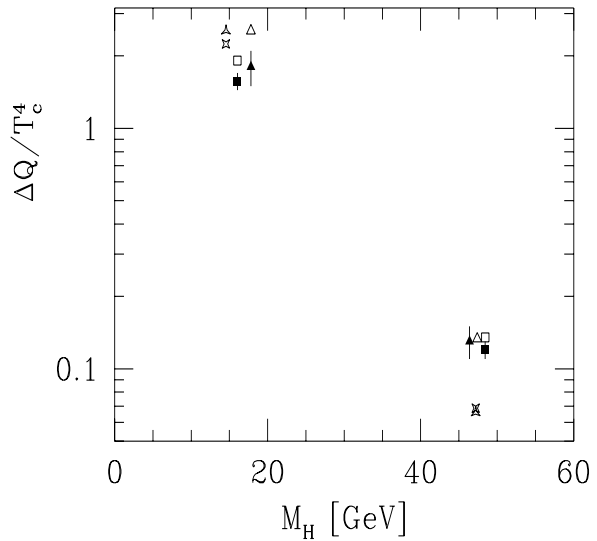

Figure 4. Latent heat as function of the zero temperature Higgs mass $M_H$. The full symbols are the definition using the derivative of the action density with respect to the lattice spacing [40]. The open symbols use the definition by the Clausius-Clapeyron equation (16). The diamond shaped symbols are from lattice perturbation theory. Squared symbols denote $L_t = 3$, triangles $L_t = 2$.

In Fig. 4 a comparison is made for both definitions of the latent heat. In the same figure results from lattice perturbation theory are plotted. There are several important features worth mentioning in the graph. First one notices that for $M_H \approx 20$GeV both definitions and the results from lattice perturbation theory agree very well. At the higher Higgs mass again the two definitions coincide but lattice perturbation theory does not describe the data as well any more. Second, there is a rapid decrease of the values for the latent heat when the Higgs mass is increased indicating that the phase transition becomes rapidly weaker. Finally, there is almost no difference between the $L_t = 2$ and the $L_t = 3$ data. The scaling violations seem to be very small which will be confirmed by other quantities later again.

### 4.2. Surface tension

Another important quantity characterizing a first order phase transition is the surface tension. Three independent methods to compute the surface tension will be shortly described and compared in this section. The first utilizes distributions of suitable observables like the action density. The distributions will develop a two peak structure at a first order phase transition. Of great help to obtain the distributions is the multicanonical simulation technique with reweighing [44,49]. If by $P_{max}$ and $P_{min}$ the maximum and the minimum of the distributions are denoted, then a finite volume estimate of the surface tension is

$$\sigma_V = \frac{1}{2L_x L_y L_t} \log \frac{P_{max}}{P_{min}} . \quad (19)$$

Taking lattices elongated in one direction, chosen to be $L_z$ and $L = L_x = L_y$, the infinite volume value for the surface tension is given by [50]

$$\sigma_\infty = \sigma_V + \frac{1}{L^2 L_t}(c + \frac{3}{4} \log L_z - \frac{1}{2} \log L) \quad (20)$$

where $c$ is an unknown constant that can be fitted by choosing different size lattices.

As a second method the above mentioned $2 - \kappa$ method [51] can be used. For the SU(2)-Higgs model it amounts to measure the link expectation values $L_\varphi = <\frac{1}{2}\text{Tr}\varphi_x^\dagger U_{x,\mu}\varphi_{x+\mu}>$ in the two halfs of the lattice with different $\kappa_1 < \kappa_c$ and $\kappa_2 > \kappa_c$. Choosing again in the $z$-direction elongated lattices one obtains [40]

$$\sigma = L_z|\kappa_{1,2} - \kappa_c|(L_\varphi^1 - L_\varphi^2) . \quad (21)$$

A third method is measuring appropriate correlation functions [52–54,36,42]. When the systems tunnels between the two states at the first order phase transition, the lowest excitation of the transfer matrix spectrum is the so called tunnel energy $E_0$. It is directly connected to the surface tension by its exponential finite size effect

$$E_0 = C \exp\{-L_t L_x L_y \sigma\} . \quad (22)$$

This method was recently applied to the SU(2)-Higgs model in [36,42].

Table 2
Comparison of the surface tension $\sigma$ as obtained from the three methods described in the text. All results are obtained for a $L_t = 2$ lattice.

| $M_H$ [GeV] | $\sigma/T_c^3$ | | |
| --- | --- | --- | --- |
| | $2 - \kappa$ | distribution | tunneling |
| 18 | 0.84(16) | 0.83(4) | -- |
| 35 | 0.065(10) | -- | 0.053(5) |
| 49 | 0.008(2) | -- | -- |

Table 2 shows that the different methods give compatible results. Since these methods are completely independent from each other, the agreement strengthens the trust in the numbers obtained from the numerical simulations. Note that also for the surface tension a rapid decrease of the strength of the phase transition is observed. Taking numbers for lattices with larger $L_t$ almost no scaling violation effects can be seen [40]. It should be noted that for $M_H \lesssim 35$GeV, $\sigma/T_c^3$ agrees amazingly well with perturbatively computed values [39].

### 5. Comparison to Perturbation theory

In the previous two sections details of the numerical simulations in the 3-dimensional reduced and in the 4-dimensional theory were given. For both approaches it was outlined how the continuum limit is reached. As fig. 3 indicates, in the 3-dimensional reduced theory a linear extrapolation of physical observables in $\beta_G$ and hence in the lattice spacing $a$ lead to continuum values. Quadratic forms of the extrapolation do not change the results much. In the 4-dimensional model the smallness of the scaling violations, which are compatible to the size of the statistical errors, lead to trustworthy continuum estimates. Since also finite size effects are under control in both approaches, the results are comparable to continuum computations like 2-loop perturbation theory.

Before a real comparison can be made, the coupling parameters used in the different approaches should be matched. As a reference point, the parameter values in the 4-dimensional theory, $M_W = 80 \text{GeV}$ and $g_R^2 \approx 0.57$ [40], are taken. The results from perturbation theory can then be related to the simulation results by choosing these values as input for the 2-loop formulae. It is important to choose also a renormalization scheme which is close to the lattice scheme for the comparison [3]. The renormalization scheme dependence can have substantial effects. To relate the data from the 3-dimensional simulations to the 4-dimensional ones, the best the author found was an extrapolation for physical quantities to the parameters of the 4-dimensional simulation using 2-loop formulae. The physical values for the SU(2)-Higgs model in the 3-dimensional reduced model are given in [26].

A first important quantity is, of course, the transition temperature itself. The simulation data for the quantity $T_c/M_H$ are very well described by lattice perturbation theory [40]. It is therefore justified to perform the continuum extrapolation $L_t \to \infty$ by using lattice perturbation theory. To get some error estimate, the error from the Monte Carlo data is taken for the extrapolated value. The situation is shown in fig. 5 [3]. The two solid curves are the continuum 2-loop perturbation theory results [5], corresponding to the two renormalized gauge couplings found numerically at the corresponding two Higgs mass values. In fig. 5 (and fig. 6) the error of the results from perturbation theory coming from higher orders is left out. An estimate of this error can be given by taking the difference between the 1-loop and 2-loop results (see figs. 4 and 5 in [5]). It turns out that the discrepancy between the numerical simulation and the perturbatively computed results is of the same order as this error for the physical quantities considered in this section. Figure 8 suggests that the lattice data are consistent with the ones from perturbation theory, although the simulation data have the tendency to be below the curves from perturbation theory. The same

---
[3] At $M_H = 35$ GeV only the value of $T_c/M_H$ from lattice perturbation theory alone is given, assuming the same size of the error as at the other Higgs mass values

is true for the situation in the 3-dimensional reduced model, see table 1.

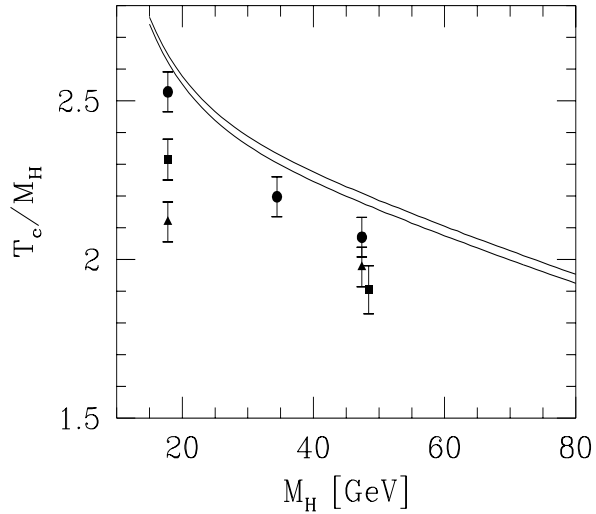

Figure 5. $T_c/M_H$ as a function of $M_H$. The solid lines are from continuum 2-loop computations [5]. The full symbols are from the 4-dimensional simulation with square denoting $L_t = 3$ and triangles $L_t = 2$ results. The full circles are results from lattice perturbation theory extrapolated to the continuum and the error taken from the simulation data.

In fig. 6 the situation for the important quantity $v_T/T_c$ is given. The scalar expectation value $v_T$ is in each case computed from the jump $\Delta$ of the gauge invariant quantity $\rho_x^2$, $v_T = \sqrt{\Delta <\rho_x^2>}$. Here lattice perturbation theory does not describe the data as well as for $T_c/M_H$ (see fig. 4) and is therefore left out. In fig. 6 results from simulations in the 3-dimensional reduced model are included, denoted by the starred symbols. It can be seen that the different lattice simulation methods are compatible. In addition, although the perturbatively obtained curves do not coincide with the Monte Carlo data, the agreement is surprisingly good, taken the general believe that perturbation theory might be inadequate to describe the electroweak phase transition.

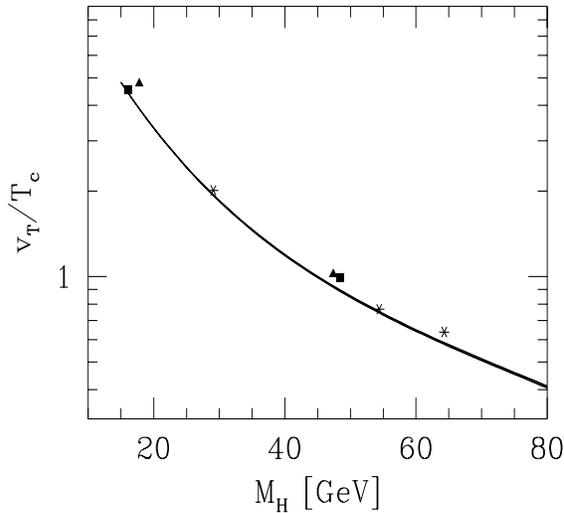

Figure 6. Same as fig. 5 for $v_T/T_c$. In this figure the results from the simulations in the 3-dimensional reduced model are included, denoted by starred symbols.

## 6. Some open questions

### 6.1. symmetric phase

Although the results as summarized in the previous section suggests that perturbation theory describes the electroweak phase transition reasonably well, the situation for the physics of the high temperature symmetry restored phase is not as clear. In [55] an attempt was made to compute the spectrum in the symmetric phase within the 3-dimensional SU(2)-Higgs model. Starting with gap-equations, the phase transition itself and the behaviour of several observables close to the transition could be described self-consistently. In particular, it was found that in the symmetric phase, the effective model describing the system is again a SU(2)-Higgs model with the form of eq.(3). Only the values of the couplings are different in the symmetric and the broken phases. This appealing picture suggests that also the symmetric region can be treated perturbatively. If both regions are really analytically connected as indicated in fig. 1, the above scenario appears to be very plausible.

The results from the work in [55] can be tested partly by Monte Carlo simulations. For example, the spectrum in the broken phase at finite temperature agrees very well between perturbation theory and numerical simulations. However, the situation in the symmetric phase is drastically different. Here the quantities of interest are masses compared to the confinement scale $g^2 T$. In [55] an upper bound for $m_W/g^2 T$ with $m_W$ the vector boson mass was given. The vector boson mass can also be determined in numerical simulations from suitable correlation functions. The results are shown in fig. 7. Here the values from simulations in 4 dimensions [40] and a recent simulation in the 3-dimensional Higgs model at $M_H = 35$ GeV [56] are shown. The shaded area gives the allowed values from [55] with the upper bound for $m_W/g^2 T < 0.29$. A clear discrepancy is encountered. Simulation results in the effective 3-dimensional reduced model [27] are compatible with the numbers from the other two simulations as shown in the plot.

At the moment it is not clear what the reason for this large difference is. The problem in the numerical simulations is that the correlation functions are very noisy in the symmetric phase and one might easily miss a low lying state as it disappears in the noise of the tail of the correlation functions. On the other hand, the picture as developed and described above in perturbation theory may not be valid. Clearly, more analytical and numerical work has to be done to clarify this important question.

### 6.2. sphaleron rates

For an estimate of the rate of baryon number violating processes in the minimal standard model a knowledge of the sphaleron transition rate is very important. An earlier review of the work done on the sphaleron transition rate with emphasis on numerical simulations and with a much more comprehensive list of references can be found in [12]. In the low temperature phase the sphaleron rate describes how fast the baryon asymmetry is washed out after completion of the phase transition. Existing results for the sphaleron rate come from the semi-classical approximation [10] and a non-perturbative test of these results is highly desirable. For this, how-

time and

$$\Delta(t) = \frac{1}{Z} \int \mathcal{D}p \mathcal{D}q \Delta C(t) e^{-H(p,q)/T} \qquad (23)$$

with

$$\Delta C(t) = [C(p(t), q(t)) - C(p(0), q(0))]^2 \qquad (24)$$

is the canonical ensemble average at real time $t$ with respect to the classical ensemble defined by the Hamiltonian $H$.

For the low temperature regime of the 2-dimensional U(1)-Higgs model the semi-classical result has been worked out exactly [58]. The data from numerical simulations agree completely with the semi-classical ones. This is illustrated in fig. 8. The figure shows $\ln F$ with $F = \Gamma/m_\Phi^2 L$ and $m_\Phi$ the scalar mass in the U(1)-Higgs model. $\ln F$ is plotted against the rescaled inverse temperature $\beta'$ (see [57]). The figure suggests that at least for low temperatures the classical description of the system is valid. This gives confidence in the semi-classical estimate for the transition rate shortly after the phase transition was completed. It would be nice to see that the same is true for the 4-dimensional SU(2)-Higgs model and therefore to confirm eq.(1) which leads to relation (2).

At high temperatures where the system is in the symmetric phase, no exact analytical results are available. In fig. 8 the numerically obtained results are fitted to a $T^2$ behaviour. The data seem to respect this form, see however [59]. The two different curves correspond to two values of the lattice spacing indicating that the rate is lattice spacing dependent. The results as obtained in the 2-dimensional model are very promising. It seems to be possible in the near future that also classical simulations in the 4-dimensional SU(2)-Higgs model can be performed with the help of the method suggested in [60]. In 4 dimensions the sphaleron transition rate is expected to show a $T^4$ law, $\Gamma = k(\alpha_W T)^4$. Recent results indicate that for the pure gauge theory in four dimensions the $T^4$ behaviour of the rate is realized with a *classical* value of the coefficient $k = 1.09(4)$ [61]. A recent perturbative estimate in the 3-dimensional SU(2)-Higgs model reveals, however, $k \approx 0.01$ [62]. It remains to be seen, whether the inclu-ever, one would need a real time simulation of the full quantum system. Unfortunately, such a method, up to now, does not exist. The best that can be done are real time simulations of the classical system using *classical* Hamilton equations of motion.

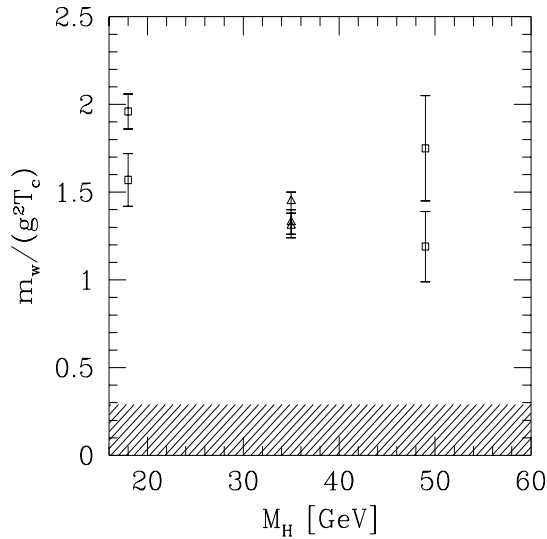

Figure 7. The finite temperature vector boson mass divided by the confinement scale $g^2T$. The shaded area indicates the allowed value as given in [55]. The simulation points are taken from [40] and from [56].

These classical real time simulations are performed by taking a Langevin type of algorithm to produce thermalized configurations [57]. One may alternatively use realistic heat bath methods [59]. The thermalized configurations are then taken as input for the integration of classical Hamilton equations of motion. It is very important that in the simulations the Gauss constraints are respected. For the 2-dimensional abelian Higgs model, which has been investigated mostly so far, this is easy to implement. However, it was only recently that a method was found which respects the Gauss constraints exactly also for non-abelian gauge theories with or without matter fields [60].

The sphaleron rate $\Gamma$ is related to the change $\Delta$ of the Chern-Simons number $C$ during the real time simulation, $\Delta(t) = \Gamma t$ where $t$ denotes real

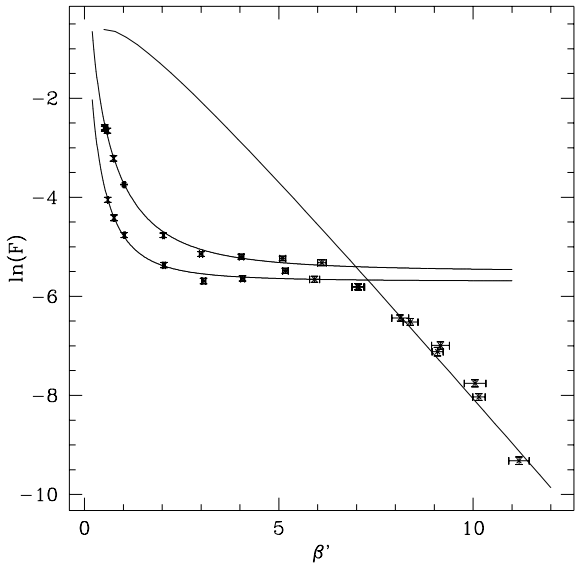

Figure 8. The sphaleron rate as a function of the inverse scaled temperature $\beta'$ [57]. For low temperatures the solid line is the exact result [58]. The curves at high temperature are fits, assuming a $1/\beta'^2$ behaviour.

sion of matter fields can change the coefficient $k$ that drastically.

### 6.3. crossover scenario

It has been suggested by various groups doing numerical simulations [16], perturbation theory [55] and average action approach [34] that for a large enough Higgs mass $M_H \approx 100$ GeV the first order electroweak phase transition changes into an analytical crossover behaviour. It would be very important if one could clarify this picture. One way would be to repeat the old simulations which were done at -for todays standards- rather small lattices. An alternative would be to find an analyticity proof similar to the Osterwalder-Seiler proof [63] for the zero temperature system.

### 6.4. linked cluster expansion

As already mentioned in the text, the dimensional reduction program, performed on the level of the action, has been criticized. It would be important to clarify the role of the non-local terms appearing when the reduction is done at the 2-loop level. Approaches to shed light on this problem or circumventing it are the concept of block spin transformation [32] or the matching of Green's functions [26]. Another interesting method is a recently performed high order linked cluster –hopping parameter– expansion [64]. The computation was performed up to the 18th order in N-component scalar field theories. As a first output, the method allows for a precise determination of critical exponents.

As a result, it could be shown that the $\phi^4$ model with $N = 4$ is governed by the 3-dimensional critical exponents at high temperatures. This is in agreement with earlier lattice simulation results [17,18] and analytical investigations [19]. In addition, it was found that the renormalized quartic coupling is weak close to the critical temperature and that $\phi^6$ terms might be important.

## 7. Conclusions

In the introduction a moderate aim was formulated, namely, to confront results as obtained in perturbation theory with data from non-perturbative lattice simulations for Higgs masses $M_H \lesssim 70$GeV. This goal has certainly been achieved by both, the 4-dimensional and the 3-dimensional reduced approaches. These two methods of investigating the electroweak phase transition by means of numerical computations give compatible results providing an important cross check of the numerical data. The most surprising outcome from the numerical simulations is that up to Higgs masses of about 70GeV the phase transition is well described by 2-loop perturbation theory. This is in contrast to earlier expectations that perturbation theory will not be able at all to say anything about the transition when physics of the symmetric phase has to be taken into account. It is a safe conclusion that we have a qualitative and quantitative good understanding of the finite temperature electroweak phase transition up to Higgs masses of about 70GeV.

It is now well established that the electroweak phase transition is of first order if the Higgs mass is taken to be smaller than about 70GeV. The strength of the phase transition decreases rapidly with increasing Higgs mass. This is seen in basically all physical observables, latent heat, sur-face tension and jump of the scalar field expectation value. The decrease is so pronounced that it becomes hard to imagine that for realistic Higgs mass values which are not ruled out by experiment the electroweak phase transition can produce the right amount of baryon asymmetry. In order to complete the picture, simulations at a Higgs mass value of about the physical W-boson mass are mandatory. However, there the phase transition might be so weak that it will become very difficult to resolve its order [27].

Although, as mentioned above, the numerically obtained results are in reasonable agreement with the ones obtained in perturbation theory, there are slight discrepancies. In these differences the non-perturbative effects are presumably hidden. It seems that for an accurate quantification of the non-perturbative effects very precise numbers from numerical simulations are necessary. For this purpose computations in the 3-dimensional reduced model are probably the most appropriate.

Open questions concern the high temperature regime. It is still not clear whether the phase transition will turn into the anticipated analytical crossover behaviour at Higgs masses at about 100GeV. The physics of the symmetric phase has to be clarified. This includes on the one hand the spectrum where a clear discrepancy between lattice simulations and perturbative calculations are encountered. On the other hand there is still a lot of work left to obtain a better understanding of the sphaleron rate.

**Acknowledgement**

It is a pleasure to thank my collaborators F. Csikor, Z. Fodor, J. Hein, A. Jaster and I. Montvay for many useful discussions. K. Farakos, K. Kajantie, M. Laine, K. Rummukainen and M. Shaposhnikov are gratefully acknowledged not only for very interesting and helpful communications but also for making their results available to me prior to publication. I thank A. Hebecker for his help in relating the different approaches to each other. I am indebted to J. Ambjorn, W. Buchmüller, B. Bunk and J. Smit for many helpful discussions and comments.